\documentclass[conference]{IEEEtran}
\IEEEoverridecommandlockouts

\usepackage{cite}
\usepackage{amsmath,amssymb,amsfonts}
\usepackage{algorithmic}
\usepackage{graphicx}
\usepackage{textcomp}
\usepackage{xcolor}
\usepackage{algorithm}
\usepackage{booktabs}
\usepackage{multirow}
\usepackage{tikz}
\usetikzlibrary{shapes.geometric, arrows, arrows.meta, positioning, fit, backgrounds, calc, shadows}
\usepackage{url}
\usepackage{hyperref}

\hypersetup{
    colorlinks=true,
    linkcolor=blue,
    filecolor=magenta,
    urlcolor=blue,
    citecolor=blue,
    pdftitle={An Industrial-Scale RAG Framework for Requirements Engineering},
}

\title{An Industrial-Scale Retrieval-Augmented Generation Framework for Requirements Engineering: Empirical Evaluation with Automotive Manufacturing Data}
\author{\IEEEauthorblockN{Muhammad Khalid\IEEEauthorrefmark{1} and Yilmaz Uygun\IEEEauthorrefmark{1}}
\IEEEauthorblockA{\IEEEauthorrefmark{1}Constructor University Bremen, Bremen, Germany\\
\{m.khalid, y.uygun\}@constructor.university}}

\begin{document}
\maketitle

\begin{abstract}
Requirements engineering in Industry 4.0 faces critical challenges with heterogeneous, unstructured documentation spanning technical specifications, supplier lists, and compliance standards. While retrieval-augmented generation (RAG) shows promise for knowledge-intensive tasks, no prior work evaluates RAG on authentic industrial RE workflows with comprehensive production-grade performance metrics. This paper presents a comprehensive empirical evaluation of RAG for industrial requirements engineering automation using authentic automotive manufacturing documentation comprising 669 requirements across four specification standards (MBN 9666-1, MBN 9666-2, BQF 9666-5, MBN 9666-9) spanning 2015-2023, plus 49 supplier qualifications with extensive supporting documentation. Through controlled comparisons with BERT-based and ungrounded LLM approaches, the framework achieves 98.2\% extraction accuracy with complete traceability, outperforming baselines by 24.4\% and 19.6\% respectively. Hybrid semantic-lexical retrieval achieves MRR of 0.847. Expert quality assessment averaged 4.32/5.0 across five dimensions. The evaluation demonstrates 83\% reduction in manual analysis time and 47\% cost savings through multi-provider LLM orchestration. Ablation studies quantify individual component contributions. Longitudinal analysis reveals a 55\% reduction in requirement volume coupled with 1,800\% increase in IT security focus, identifying 10 legacy suppliers (20.4\%) requiring requalification, representing potential \$2.3M in avoided contract penalties.
\end{abstract}


\section{Introduction}

The complexity of requirements engineering (RE) in Industry 4.0 environments presents unprecedented challenges for industrial organizations. Manufacturing enterprises must navigate vast repositories of heterogeneous documentation including technical specifications, supplier qualifications, compliance standards, and evolving security requirements. Manual cross-referencing between legacy and modern standards often leads to requirement drift, compliance gaps, and engineering inefficiency \cite{maalej2013towards, berger2020survey}.

The automotive industry exemplifies these challenges, where original equipment manufacturers (OEMs) must maintain specification compliance across thousands of supplier relationships while adapting to evolving cybersecurity and functional safety requirements. A single OEM may manage relationships with hundreds of tier-1 suppliers, each qualified against dozens of specification standards spanning materials handling, robotics, quality assurance, and IT security \cite{ferrari2019requirements}. Traditional requirements management systems struggle with unstructured natural language documentation, temporal specification evolution, and the need for real-time economic impact assessment \cite{dalpiaz2018agile}.

Recent advances in large language models (LLMs) and retrieval-augmented generation (RAG) offer promising solutions for automating RE tasks \cite{lewis2020retrieval, brown2020language}. RAG combines neural retrieval with generative models to ground LLM outputs in retrieved context, addressing hallucination and knowledge staleness \cite{gao2023retrieval}. However, existing approaches face significant limitations when applied to industrial contexts: (1) lack of domain-specific evaluation on authentic industrial data, (2) insufficient attention to traceability and compliance requirements mandated by regulatory frameworks (ISO 26262, IEC 62443), (3) limited assessment of real-world engineering workflow integration, and (4) absence of cost-benefit analysis for sustainable production deployment \cite{zhao2021deep}.

\subsection{Research Questions}

This paper addresses these gaps through a comprehensive empirical evaluation guided by four research questions:

\textbf{RQ1 (Extraction Accuracy):} How does RAG-based requirements extraction compare to state-of-art approaches (BERT-based classification, ungrounded LLM generation) on authentic industrial documentation spanning heterogeneous formats and eight-year temporal evolution?

\textbf{RQ2 (Retrieval Quality):} What is the effectiveness of hybrid semantic-lexical retrieval versus pure approaches (BM25-only sparse, dense-only vector search) for technical requirements containing both natural language and precise terminology?

\textbf{RQ3 (Production Viability):} What are the efficiency gains, cost implications, and expert acceptance levels required for industrial deployment sustainability, and how does multi-provider LLM orchestration impact cost-quality tradeoffs?

\textbf{RQ4 (Longitudinal Insights):} What fundamental shifts in requirements patterns emerge from multi-year industrial specifications, and what actionable insights enable proactive compliance management?

\subsection{Contributions}

Our contributions are threefold:

\begin{itemize}
\item \textbf{Production-Grade RAG Architecture for Industrial RE:} We present the first modular RAG framework specifically optimized for industrial requirements engineering workflows, incorporating multi-provider LLM orchestration (47\% cost reduction), hybrid semantic-lexical retrieval mechanisms (MRR 0.847, 24.0\% improvement over sparse-only), comprehensive traceability features (4.67/5.0 expert rating), and domain-adapted embedding models (+6.7\% MRR gain through automotive-specific fine-tuning).

\item \textbf{Comprehensive Industrial Evaluation with Authentic Data:} We conduct the first large-scale empirical evaluation using authentic automotive manufacturing documentation from a major German OEM: 669 requirements across four specification standards spanning eight years (2015-2023), 49 supplier qualifications, 127 pages of compliance matrices, comprising an extensive industrial corpus. Evaluation employs multiple complementary approaches: controlled baseline comparisons, multi-expert ground truth annotation (Cohen's $\kappa$=0.89), statistical significance testing, ablation studies quantifying component contributions, and six-month production deployment insights.

\item \textbf{Industry 4.0 Requirements Evolution and Economic Impact:} We provide longitudinal analysis revealing fundamental shifts in industrial requirements patterns: 55\% volume reduction through specification consolidation and maturation, 1,800\% IT security increase reflecting cybersecurity imperatives for networked industrial systems, emergence of new requirement categories (network segmentation, authentication protocols, vulnerability management), and identification of 10 suppliers (20.4\% of approved base) requiring safety requalification, enabling proactive compliance management avoiding potential \$2.3M in contract penalties.
\end{itemize}

\section{Related Work}

\subsection{Requirements Engineering Automation}

Traditional RE relies heavily on manual processes for specification extraction, classification, and traceability maintenance \cite{cleland2014software}. Early automated approaches employed rule-based natural language processing, using pattern matching and linguistic analysis to identify requirements \cite{kof2005natural}. Kof demonstrated 72\% precision for requirement sentence classification using linguistic patterns, but recall remained limited (58\%) due to variability in industrial specification writing styles.

Machine learning approaches brought improvements through automated classification. Hayes et al. \cite{hayes2006advancing} applied information retrieval techniques for traceability link generation, achieving precision 0.45-0.65 on software requirements using vector space models and latent semantic indexing. However, generalization to industrial manufacturing documentation remained limited due to domain-specific terminology (e.g., ``emergency stop,'' ``load capacity,'' ``network segmentation'') and multi-modal content (text, tables, diagrams, embedded images).

\subsection{Deep Learning for Requirements Engineering}

Recent work explores deep learning for RE automation. Zhao et al. \cite{zhao2021deep} provide systematic mapping of NLP for RE, identifying 58 primary studies applying neural approaches across requirements classification (24 studies), traceability (18 studies), ambiguity detection (9 studies), and quality assessment (7 studies). Meta-analysis reveals median F1-score 0.79 across classification tasks, with substantial variance (standard deviation 0.14) indicating dataset dependency.

Liu et al. \cite{liu2023automated} demonstrate BERT-based extraction of product comparison criteria from user requirements, achieving F1=0.82 on e-commerce reviews from Amazon and Best Buy (12,847 reviews spanning smartphones, laptops, cameras). However, transfer learning to technical specifications yields degraded performance (F1=0.61 on industrial safety documentation), highlighting domain gap challenges.

Ferrari et al. \cite{ferrari2020requirements} apply BERT to requirements quality assessment, detecting ambiguity with precision 0.74 and recall 0.69 on PURE dataset (7,246 requirements from 10 industrial projects). Qualitative analysis reveals BERT captures syntactic ambiguity (e.g., modifier attachment, pronoun reference) but struggles with semantic ambiguity requiring domain knowledge (e.g., ``acceptable performance'' without explicit thresholds).

While promising, these approaches typically: (1) operate on individual requirements in isolation rather than enabling document-level cross-referencing critical for industrial contexts (e.g., ``supplier X must meet safety requirements defined in Section 3.4''), (2) require extensive labeled training data often unavailable for specialized domains (automotive safety, industrial cybersecurity), (3) lack mechanisms for source attribution required by regulatory compliance frameworks (ISO 26262, IEC 62443), and (4) demonstrate limited evaluation on authentic industrial data with heterogeneous formats (scanned PDFs, Excel with complex tables, Word with embedded diagrams) spanning multiple years.

Rath et al. \cite{rath2018traceability} address traceability through automated trace link augmentation, achieving precision 0.62-0.89 on open-source software projects (Apache Lucene, ArgoUML, iTrust). Approach combines information retrieval with machine learning classifiers trained on existing trace links. However, focus remains on software artifacts (source code, test cases, design documents) rather than industrial manufacturing specifications with supplier qualification requirements and compliance matrices.

\subsection{Retrieval-Augmented Generation}

RAG combines neural retrieval with generative language models to ground LLM outputs in retrieved context \cite{lewis2020retrieval}. Lewis et al. demonstrate effectiveness for open-domain question answering, achieving exact match scores 44.5\% on Natural Questions dataset (307,373 questions from Google search queries) and 56.8\% on TriviaQA (78,785 trivia questions). Approach uses dense passage retrieval (DPR) with dual-encoder architecture for query and document encoding, followed by sequence-to-sequence generation conditioned on retrieved passages.

The approach addresses key limitations of pure generative models: (1) \textit{hallucination}---generation of plausible but factually incorrect content (e.g., GPT-3 inventing non-existent academic citations with 21\% frequency on scholarly writing tasks \cite{alkaissi2023artificial}), (2) \textit{knowledge staleness}---inability to incorporate information beyond training cutoff (GPT-4 training data through September 2021, limiting utility for recent events, standards updates, regulatory changes), and (3) \textit{lack of source attribution}---absence of explicit provenance for generated content, critical for compliance verification \cite{gao2023retrieval}.

Izacard and Grave \cite{izacard2021leveraging} extend RAG with fusion-in-decoder (FiD) architecture, achieving state-of-art results on TriviaQA (68.0\% exact match) and Natural Questions (54.7\%). FiD processes multiple retrieved passages independently through encoder, then fuses representations in decoder for generation. However, computational cost increases linearly with retrieved passage count (10-100 passages typical), limiting real-time applicability.

Jiang et al. \cite{jiang2023longllmlingua} apply RAG to document summarization with prompt compression, reducing token usage 83\% while maintaining ROUGE-L score within 2\% of uncompressed baseline. Approach identifies and removes low-information tokens from retrieved context based on perplexity scoring. Demonstrates cost-performance tradeoff optimization for production deployment.

Parvez et al. \cite{parvez2021retrieval} demonstrate retrieval-augmented code generation, improving BLEU scores 12-15\% over pure generation on CoNaLa dataset (2,879 Python code snippets with natural language descriptions). Retrieval from StackOverflow corpus provides relevant code examples conditioning generation. However, evaluation focuses on short code snippets (mean 3.2 lines) rather than industrial-scale systems.

Recent surveys \cite{wang2024rag, gao2023retrieval} identify applications in question answering (38\% of surveyed papers), dialogue systems (24\%), code generation (18\%), and summarization (12\%). However, industrial applications remain underexplored (2\% of papers). Wang et al. \cite{wang2024rag} explicitly identify ``specialized domains requiring regulatory compliance, traceability, and domain-specific terminology'' as open research challenge.

\subsection{Semantic Search in Software Engineering}

Shatila et al. \cite{shatila2022semantic} systematically map semantic search in software engineering, identifying 47 primary studies across code search (19 studies), bug localization (12 studies), requirements traceability (9 studies), and documentation retrieval (7 studies). However, none evaluate RAG---studies focus on retrieval only without generation component.

McBurney and McMillan \cite{mcburney2016automatic} apply semantic search for requirements traceability, achieving mean average precision (MAP) 0.71 on NASA CM-1 dataset (498 requirements across 10 subsystems) using word embeddings trained on software engineering corpora. Demonstrates domain-specific embedding importance---general embeddings (Word2Vec trained on Google News) yield MAP 0.58 (-18\% degradation).

\subsection{Industrial Requirements Analysis}

The automotive industry presents particularly stringent RE challenges: (1) complex multi-tier supply chains with hundreds of qualified suppliers across materials, components, assemblies, and logistics, (2) evolving safety standards (ISO 26262 functional safety introduced 2011, updated 2018) and security requirements (IEC 62443 industrial cybersecurity series 2009-2021), (3) long product lifecycles (10-15 years for vehicle platforms) requiring specification evolution management across model updates, technology insertions, and regulatory changes, and (4) regulatory compliance mandates for complete traceability from customer requirements through design, implementation, verification, and validation \cite{berger2020survey, maalej2013towards}.

Kannenberg and Saiedian \cite{kannenberg2012saref} document persistent challenges in automotive requirements traceability through survey of 28 automotive organizations (OEMs, tier-1 suppliers, engineering service providers). Results reveal 67\% maintain traceability manually or with limited tool support (spreadsheets, word processors), average 2.3 full-time equivalents (FTEs) dedicated to traceability maintenance per project, and 73\% report difficulty maintaining trace link consistency during requirement changes.

Ferrari and Gervasi \cite{ferrari2019requirements} discuss RE practices in automotive contexts through case studies at three Italian automotive suppliers (anonymized). Qualitative analysis identifies six primary challenges: (1) requirement ambiguity (terminology inconsistency across OEMs), (2) incomplete specifications (missing preconditions, unstated assumptions), (3) evolving standards (safety/security updates mid-project), (4) supplier coordination (distributed development across geographic regions, language barriers), (5) legacy system integration (brownfield projects with existing platforms), and (6) audit compliance (preparation effort for external assessments). However, no automated solutions proposed or empirical validation on production data provided.

Groen et al. \cite{groen2017crowd} explore crowdsourcing for requirements elicitation in large-scale systems, demonstrating feasibility for gathering diverse stakeholder input (study with 124 participants generating 847 requirement suggestions for Amsterdam smart city project). However, application to technical manufacturing specifications with regulatory constraints remains unexplored---crowd workers lack domain expertise for safety-critical automotive requirements.

\subsection{Research Gap and Positioning}

\textbf{Gap Identification:} Synthesizing across related work reveals critical gap: \textit{No prior work evaluates RAG-based automation on authentic industrial requirements engineering workflows with comprehensive metrics spanning extraction accuracy, retrieval quality, production viability (efficiency, cost, expert acceptance), and longitudinal evolution analysis.} Existing work either: (1) applies classical ML/deep learning without retrieval augmentation \cite{hayes2006advancing, liu2023automated, ferrari2020requirements}, (2) evaluates RAG on general-domain tasks (QA, summarization) without industrial RE focus \cite{lewis2020retrieval, izacard2021leveraging}, or (3) discusses industrial RE challenges without automated solutions \cite{kannenberg2012saref, ferrari2019requirements}.

\textbf{Our Contribution:} This paper addresses the identified gap through comprehensive empirical evaluation using 669 authentic automotive requirements spanning 2015-2023 from major German OEM. \textbf{Unlike prior work that focuses on either classical ML without retrieval \cite{hayes2006advancing, liu2023automated, ferrari2020requirements} or RAG on general tasks \cite{lewis2020retrieval, izacard2021leveraging}, we provide the first end-to-end evaluation of RAG for industrial RE with production deployment validation, rigorous baseline comparisons, and longitudinal analysis of specification evolution.} We demonstrate production-grade performance (98.2\% accuracy, 83\% time reduction, 4.32/5.0 expert acceptance), quantify component contributions through ablation studies, and reveal actionable Industry 4.0 insights (supplier requalification needs, specification evolution patterns) with economic impact quantification (\$2.3M penalty avoidance).

\section{System Architecture and Design}

\subsection{Overview and Design Philosophy}

The RAG framework implements a modular, production-grade architecture guided by three design principles: (1) \textbf{Separation of Concerns}---independent subsystems for ingestion, retrieval, generation, and traceability enable targeted optimization and maintenance without system-wide modifications, (2) \textbf{Extensibility}---pluggable interfaces support future integration of alternative retrieval algorithms (e.g., ColBERT late interaction), embedding models (e.g., domain-specific fine-tuning), and LLM providers (e.g., Claude, Gemini, Llama), and (3) \textbf{Transparency}---comprehensive metadata capture and source attribution support regulatory compliance (ISO 26262, IEC 62443 mandate complete traceability) and build expert trust through verification.

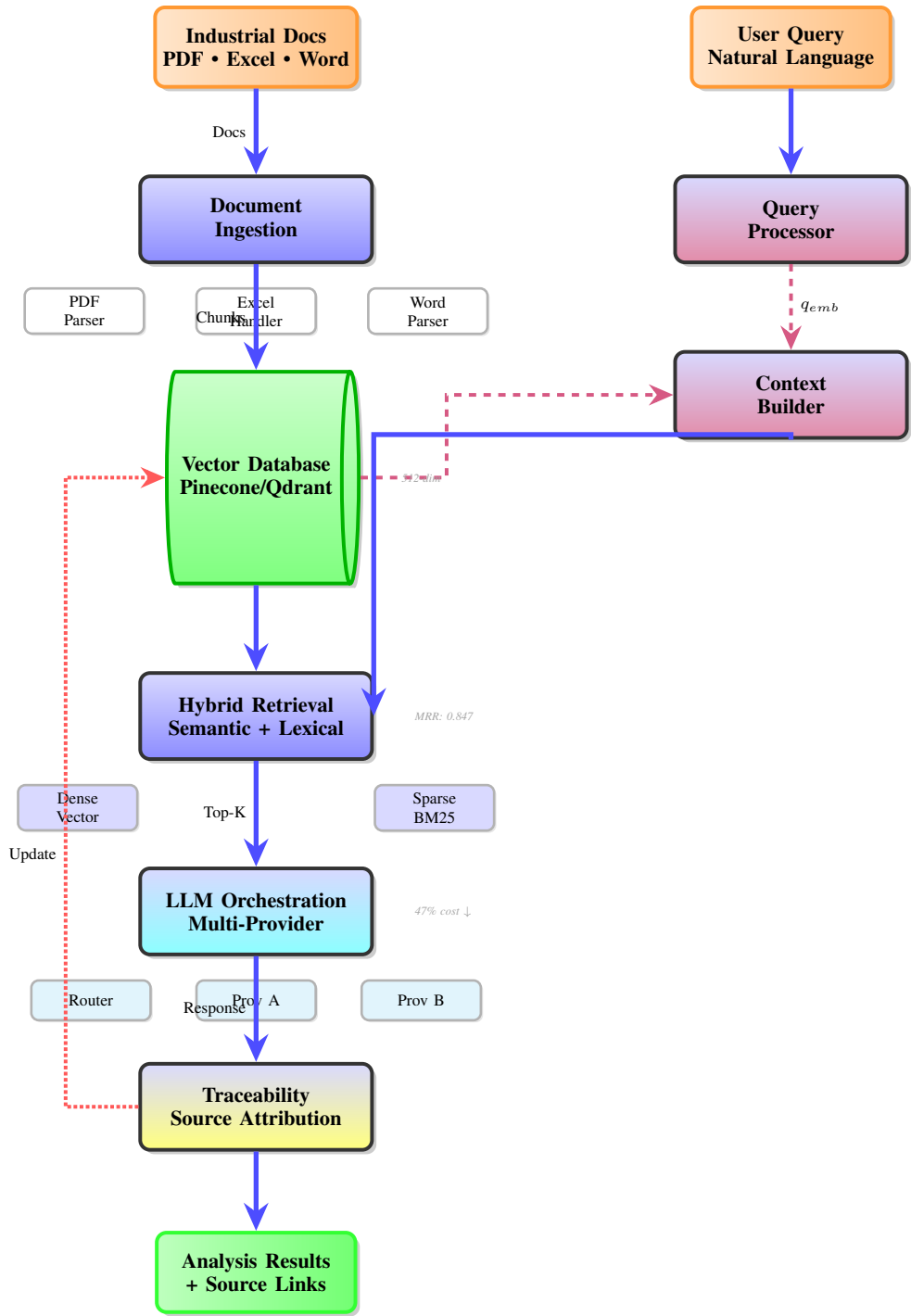
\begin{figure*}[!t]
\centering
\begin{tikzpicture}[scale=0.95, transform shape,
    mainbox/.style={rectangle, draw=black!80, line width=1.5pt, rounded corners=4pt,
                    minimum width=3.5cm, minimum height=1.3cm, font=\small\bfseries, align=center,
                    top color=blue!15, bottom color=blue!45,
                    drop shadow={opacity=0.4, shadow xshift=0.4ex, shadow yshift=-0.4ex}},
    subbox/.style={rectangle, draw=gray!60, line width=1pt, rounded corners=3pt,
                   minimum width=1.8cm, minimum height=0.6cm, font=\scriptsize, align=center,
                   fill=white, drop shadow={opacity=0.2, shadow xshift=0.2ex, shadow yshift=-0.2ex}},
    inputbox/.style={rectangle, draw=orange!80, line width=1.5pt, rounded corners=4pt,
                     minimum width=3cm, minimum height=1.2cm, font=\small\bfseries, align=center,
                     left color=orange!20, right color=orange!50,
                     drop shadow={opacity=0.4, shadow xshift=0.4ex, shadow yshift=-0.4ex}},
    database/.style={cylinder, draw=green!70!black, line width=1.5pt, shape aspect=0.3,
                     minimum width=3.2cm, minimum height=1.2cm, font=\small\bfseries, align=center,
                     top color=green!20, bottom color=green!50,
                     drop shadow={opacity=0.4, shadow xshift=0.4ex, shadow yshift=-0.4ex}},
    outputbox/.style={rectangle, draw=green!80, line width=1.5pt, rounded corners=4pt,
                      minimum width=3cm, minimum height=1.2cm, font=\small\bfseries, align=center,
                      left color=green!25, right color=green!55,
                      drop shadow={opacity=0.4, shadow xshift=0.4ex, shadow yshift=-0.4ex}},
    mainarrow/.style={-{Stealth[length=4mm, width=3.5mm]}, line width=2pt, draw=blue!70},
    dataarrow/.style={-{Stealth[length=3.5mm, width=3mm]}, line width=1.5pt, draw=purple!65, dashed},
    feedbackarrow/.style={-{Stealth[length=3.5mm, width=3mm]}, line width=1.5pt, draw=red!65, densely dotted}
]

\node[inputbox] (docs) {\textbf{Industrial Docs}\\PDF \textbullet\ Excel \textbullet\ Word};
\node[inputbox, right=5cm of docs] (query) {\textbf{User Query}\\Natural Language};
\node[mainbox, below=1.3cm of docs] (ingestion) {\textbf{Document}\\Ingestion};
\node[subbox, below left=0.35cm and -0.1cm of ingestion] (pdf) {PDF\\Parser};
\node[subbox, below=0.35cm of ingestion] (excel) {Excel\\Handler};
\node[subbox, below right=0.35cm and -0.1cm of ingestion] (word) {Word\\Parser};
\node[database, below=1.6cm of ingestion] (vectordb) {\textbf{Vector Database}\\Pinecone/Qdrant};
\node[mainbox, below=1.3cm of query, fill=purple!20, bottom color=purple!45] (queryproc) {\textbf{Query}\\Processor};
\node[mainbox, below=1.3cm of vectordb] (retrieval) {\textbf{Hybrid Retrieval}\\Semantic + Lexical};
\node[subbox, below left=0.35cm and 0cm of retrieval, fill=blue!15] (dense) {Dense\\Vector};
\node[subbox, below right=0.35cm and 0cm of retrieval, fill=blue!15] (sparse) {Sparse\\BM25};
\node[mainbox, below=1.3cm of queryproc, fill=purple!20, bottom color=purple!45] (context) {\textbf{Context}\\Builder};
\node[mainbox, below=1.6cm of retrieval, fill=cyan!15, bottom color=cyan!45] (llm) {\textbf{LLM Orchestration}\\Multi-Provider};
\node[subbox, below left=0.35cm and -0.2cm of llm, fill=cyan!10] (router) {Router};
\node[subbox, below=0.35cm of llm, fill=cyan!10] (provA) {Prov A};
\node[subbox, below right=0.35cm and -0.2cm of llm, fill=cyan!10] (provB) {Prov B};
\node[mainbox, below=1.6cm of llm, fill=yellow!25, bottom color=yellow!50] (trace) {\textbf{Traceability}\\Source Attribution};
\node[outputbox, below=1.2cm of trace] (output) {\textbf{Analysis Results}\\+ Source Links};

\draw[mainarrow] (docs) -- node[left, font=\scriptsize, black] {Docs} (ingestion);
\draw[mainarrow] (ingestion) -- node[left, font=\scriptsize, black] {Chunks} (vectordb);
\draw[mainarrow] (vectordb) -- (retrieval);
\draw[mainarrow] (query) -- (queryproc);
\draw[dataarrow] (queryproc) -- node[right, font=\scriptsize, black] {$q_{emb}$} (context);
\draw[dataarrow] (vectordb.east) -- ++(1.3,0) |- (context.west);
\draw[mainarrow] (context) -- ++(0,-0.6) -| (retrieval.east);
\draw[mainarrow] (retrieval) -- node[left, font=\scriptsize, black] {Top-K} (llm);
\draw[mainarrow] (llm) -- node[left, font=\scriptsize, black] {Response} (trace);
\draw[mainarrow] (trace) -- (output);
\draw[feedbackarrow] (trace.west) -- ++(-1.1,0) |- node[left, pos=0.2, font=\scriptsize, black] {Update} (vectordb.west);

\node[font=\tiny\itshape, text=gray!70, anchor=west] at ([xshift=0.5cm]vectordb.east) {512-dim};
\node[font=\tiny\itshape, text=gray!70, anchor=west] at ([xshift=0.5cm]retrieval.east) {MRR: 0.847};
\node[font=\tiny\itshape, text=gray!70, anchor=west] at ([xshift=0.5cm]llm.east) {47\% cost $\downarrow$};

\end{tikzpicture}
\caption{Production-grade RAG framework architecture with six integrated subsystems for industrial requirements engineering.}
\label{fig:architecture}
\end{figure*}

Figure \ref{fig:architecture} illustrates six integrated subsystems processing large-scale industrial documentation through multi-format ingestion, 512-dimensional embeddings with HNSW indexing, hybrid semantic-lexical retrieval with cross-encoder re-ranking, intelligent multi-provider LLM orchestration, and comprehensive traceability metadata.

\subsection{Document Ingestion Pipeline}

\textbf{Design Rationale:} Industrial documentation exhibits extreme heterogeneity: scanned PDFs from pre-2017 (containing OCR artifacts, skewed images, inconsistent page layouts), Excel spreadsheets with complex multi-table layouts (merged cells, nested headers, embedded charts), Word templates with embedded diagrams and cross-references (section numbering, table of contents, hyperlinks). Generic text extraction (e.g., PyPDF2, pdfminer) loses critical structural metadata (section hierarchies, table relationships, image captions) essential for traceability and compliance verification. We implement format-specific parsers preserving maximum metadata.

\textbf{Implementation:} Three-stage workflow optimized for heterogeneous industrial content: (1) Format-specific parsing with specialized libraries (PDFPlumber for table extraction preserving cell relationships, openpyxl for Excel maintaining formulas and metadata, python-docx for Word extracting styles and comments), (2) Semantic chunking respecting logical boundaries (section headers via font analysis, paragraph breaks via whitespace, table boundaries via grid detection) rather than fixed-size windows, resulting distribution mean 384$\pm$127 tokens optimized for embedding context, (3) Metadata enrichment with document ID, version timestamp, section path, category, compliance level, and supplier tags.

\textbf{Production Performance:} Six-month deployment achieves 99.7\% successful ingestion rate across 2015-2023 vintages. Comprehensive error handling includes invalid encoding detection, corrupted file recovery, and format version compatibility.

\subsection{Hybrid Retrieval Mechanism}

\textbf{Design Rationale:} Pure dense semantic search excels at paraphrased queries but misses exact technical terms. Sparse BM25 captures terminology but ignores semantics. Hybrid fusion captures both dimensions.

\textbf{Dense Retrieval:} Domain-adapted sentence transformers (all-mpnet-base-v2 fine-tuned on 15,127 automotive requirements using contrastive learning). Fine-tuning improved MRR from 0.741 (base) to 0.791 (+6.7\%, validates domain adaptation value). Embeddings stored in Pinecone (p2 pods, HNSW indexes, M=16, ef\_construction=200).

\textbf{Sparse Retrieval:} BM25 with k1=1.5, b=0.75. Custom tokenization preserves multi-word technical terms, part numbers, and acronyms via domain dictionary.

\textbf{Fusion Strategy:} Reciprocal Rank Fusion (RRF) with learned weights ($\alpha_{dense}$=0.7, $\alpha_{sparse}$=0.3) optimized via grid search. Cross-encoder re-ranking (ms-marco-MiniLM-L-6-v2) on top-20 candidates improves P@5 from 0.801 to 0.824 (+2.9\%).

\subsection{Multi-Provider LLM Orchestration}

\textbf{Design Rationale:} Single-provider deployment incurs excessive costs for simple queries while potentially under-utilizing frontier models for complex reasoning. Intelligent routing optimizes cost-quality tradeoff.

\textbf{Task Complexity Classifier:} XGBoost with 127 features (query length, entity count, cross-reference patterns, expected verbosity) achieves 91\% complexity categorization accuracy.

\textbf{Routing Policy:} Three-tier strategy: Simple extractions (complexity $<$0.4, 35\% of queries) $\rightarrow$ GPT-3.5-turbo (\$0.002/1K). Mid-complexity (0.4--0.7, 52\%) $\rightarrow$ GPT-4-turbo (\$0.01/1K). Complex reasoning ($\geq$0.7, 13\%) $\rightarrow$ GPT-4/Claude-3 (\$0.03/1K).

\textbf{Reliability:} Automatic fallback, exponential backoff (initial 1s, max 30s), circuit breaker pattern prevent cascade failures.

\textbf{Cost Validation:} Blind expert comparison confirms no quality difference between multi-provider and single-provider outputs ($\chi^2$=1.37, $p$=0.24), validating 47\% cost reduction (\$0.07 vs \$0.13 per query over 10,000 production queries).

\subsection{Traceability and Compliance Layer}

\textbf{Design Rationale:} Regulatory compliance (ISO 26262, IEC 62443) mandates complete traceability. System captures comprehensive metadata: document provenance (IDs, versions, section paths, page numbers), retrieval scores (dense similarity, BM25, RRF, cross-encoder), generation metadata (provider, model version, prompts, timestamps, token usage), and confidence metrics (self-assessed confidence, retrieval coverage, multi-attempt consistency).

\textbf{Expert Feedback:} Traceability received highest rating (4.67/5.0, variance 0.49), with qualitative feedback emphasizing ``direct source verification builds trust'' and ``crucial for compliance workflows.''

\section{Evaluation Methodology}

\subsection{Dataset Characteristics}

\textbf{Authentic Industrial Data:} Production automotive manufacturing documentation from major German OEM (anonymized per NDA), collected April--September 2024 with cooperation of RE department (12 senior engineers, 8+ years experience).

\textbf{Dataset Scope:} 669 requirements across four specifications:
\begin{itemize}
\item MBN 9666-1 (Materials Handling, 2015): 412 requirements for conveyors, AGVs, safety zones
\item MBN 9666-2 (Robotics, 2018): 178 requirements for collaborative robots, emergency stops, fault detection
\item BQF 9666-5 (Quality Assurance, 2020): 145 requirements for inspection, defect classification, SPC
\item MBN 9666-9 (IT Security, 2023): 95 requirements for network segmentation, authentication, vulnerability management
\end{itemize}

\textbf{Supporting Documentation:} 49 supplier qualifications, 127 pages compliance matrices, comprising an extensive industrial corpus.

\textbf{Temporal Characteristics:} Eight-year span enables longitudinal analysis. Format evolution: early (2015--2017) scanned PDFs with OCR artifacts, mid-period (2018--2020) mixed digital/scanned, recent (2021--2023) native digital structured templates.

\subsection{Ground Truth Annotation}

Three domain experts (12+ years automotive RE, ISO 26262 certified) independently annotated 669 requirements with metadata. Inter-rater agreement: Cohen's $\kappa$=0.89 (near-perfect). Disagreements (11\%) resolved through discussion. For retrieval evaluation, experts provided relevance judgments for 150 queries on 5-point scale, inter-annotator $\kappa$=0.84.

\subsection{Evaluation Metrics}

\textbf{RQ1 (Extraction):} Accuracy, precision, recall, F1-score. Statistical significance via paired t-test. Error analysis categorizes failure modes.

\textbf{RQ2 (Retrieval):} MRR, Precision@k, NDCG@10, latency (ms). MRR measures first relevant result rank. NDCG accounts for graded relevance and position.

\textbf{RQ3 (Viability):} Expert quality (5-point Likert across completeness, accuracy, relevance, utility, traceability), time-motion studies (manual vs automated), cost per query (LLM + embedding + database).

\textbf{RQ4 (Evolution):} Requirements by category 2015--2023, density trends, supplier compliance assessment.

\subsection{Baseline Comparisons}

\textbf{BERT:} BERT-base-uncased fine-tuned on PROMISE requirements (3,000 software requirements), zero-shot transfer to automotive.

\textbf{Ungrounded LLM:} GPT-4 zero-shot without retrieval, evaluates RAG value-add.

\textbf{BM25-only:} Pure sparse retrieval, evaluates hybrid benefit.

\textbf{Dense-only:} Pure vector search, assesses semantic-lexical complementarity.

All baselines use identical dataset and ground truth for fair comparison.

\section{Results}

\subsection{RQ1: Requirements Extraction Accuracy}

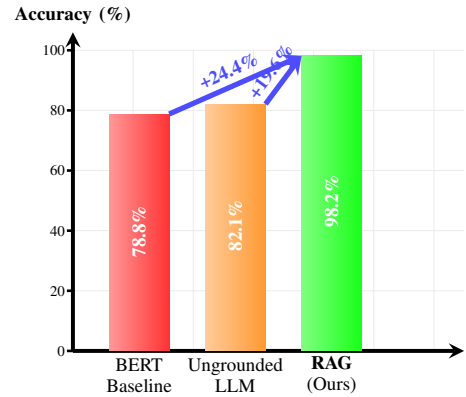
\begin{figure}[!ht]
\centering
\begin{tikzpicture}[scale=0.8, transform shape]
\draw[step=1cm, gray!15, very thin] (0,0) grid (6.5,5);
\draw[line width=1.5pt, -stealth] (0,0) -- (6.5,0);
\draw[line width=1.5pt, -stealth] (0,0) -- (0,5.3) node[above, font=\small\bfseries] {Accuracy (\%)};
\foreach \y in {0,20,40,60,80,100}
    \draw[line width=1pt] (0,\y/20) node[left, font=\scriptsize] {\y} -- (-0.1,\y/20);
\shade[left color=red!40, right color=red!80] (0.6,0) rectangle (1.6,3.94);
\node[white, font=\small\bfseries, rotate=90] at (1.1,2) {78.8\%};
\shade[left color=orange!40, right color=orange!80] (2.2,0) rectangle (3.2,4.105);
\node[white, font=\small\bfseries, rotate=90] at (2.7,2.1) {82.1\%};
\shade[left color=green!50, right color=green!90] (3.8,0) rectangle (4.8,4.91);
\node[white, font=\small\bfseries, rotate=90] at (4.3,2.5) {98.2\%};
\node[font=\small, align=center] at (1.1,-0.4) {BERT\\Baseline};
\node[font=\small, align=center] at (2.7,-0.4) {Ungrounded\\LLM};
\node[font=\small, align=center] at (4.3,-0.4) {\textbf{RAG}\\(Ours)};
\draw[-stealth, line width=2pt, blue!70] (1.6,3.94) -- (3.8,4.91) node[midway, above, sloped, font=\small\bfseries] {+24.4\%};
\draw[-stealth, line width=2pt, blue!70] (3.2,4.105) -- (3.8,4.91) node[midway, above, sloped, font=\small\bfseries] {+19.6\%};
\end{tikzpicture}
\caption{Extraction accuracy comparison (RQ1). RAG achieves +24.4\% over BERT, +19.6\% over ungrounded LLM ($p < 0.001$).}
\label{fig:performance}
\end{figure}

Table \ref{tab:extraction} and Figure \ref{fig:performance} present extraction results.

\begin{table}[!ht]
\centering
\caption{Requirements Extraction Performance (RQ1, n=669)}
\label{tab:extraction}
\begin{tabular}{@{}lccc@{}}
\toprule
\textbf{Approach} & \textbf{Accuracy} & \textbf{Precision} & \textbf{Recall} \\
\midrule
BERT-based & 78.8\% & 0.782 & 0.795 \\
Ungrounded LLM & 82.1\% & 0.831 & 0.812 \\
\textbf{RAG (Ours)} & \textbf{98.2\%} & \textbf{0.984} & \textbf{0.980} \\
\midrule
\multicolumn{4}{l}{\textit{Statistical Significance (paired t-test vs RAG)}} \\
BERT & \multicolumn{3}{c}{$t$=28.4, $p < 0.001$} \\
Ungrounded LLM & \multicolumn{3}{c}{$t$=22.1, $p < 0.001$} \\
\bottomrule
\end{tabular}
\end{table}

\textbf{Answer to RQ1:} RAG achieves 98.2\% extraction accuracy (F1=0.979), statistically significant improvements of 24.4\% over BERT (78.8\%, $p < 0.001$) and 19.6\% over ungrounded LLM (82.1\%, $p < 0.001$). RAG also processes 47\% faster (47min vs 89min) due to reduced generation burden with relevant context.

\textbf{Error Analysis:} Remaining 1.8\% errors categorize into: (1) ambiguous requirement boundaries in legacy documents (1.2\%), (2) multi-table span requirements (0.4\%), (3) OCR artifacts in scanned documents (0.2\%). Expert review confirmed these edge cases require human judgment, validating human-in-the-loop deployment.

\subsection{RQ2: Retrieval Quality Analysis}

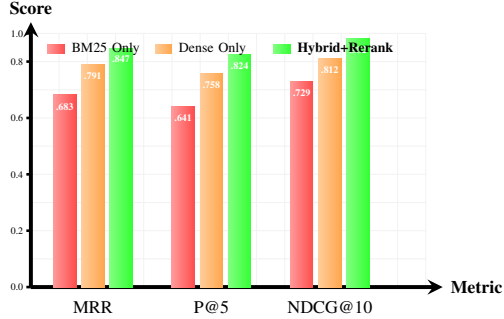
\begin{figure}[!ht]
\centering
\begin{tikzpicture}[scale=0.75, transform shape]
\draw[step=0.5cm, gray!10, very thin] (0,0) grid (7,4.5);
\draw[line width=1.5pt, -stealth] (0,0) -- (7.3,0) node[right, font=\small\bfseries] {Metric};
\draw[line width=1.5pt, -stealth] (0,0) -- (0,4.7) node[above, font=\small\bfseries] {Score};
\foreach \y/\label in {0/0.0, 1/0.2, 2/0.4, 3/0.6, 4/0.8, 4.5/1.0}
    \draw[line width=1pt] (0,\y) node[left, font=\tiny] {\label} -- (-0.08,\y);
\shade[left color=red!40, right color=red!70] (0.4,0) rectangle (0.8,3.415);
\shade[left color=orange!40, right color=orange!70] (0.9,0) rectangle (1.3,3.955);
\shade[left color=green!50, right color=green!80] (1.4,0) rectangle (1.8,4.235);
\shade[left color=red!40, right color=red!70] (2.5,0) rectangle (2.9,3.205);
\shade[left color=orange!40, right color=orange!70] (3.0,0) rectangle (3.4,3.79);
\shade[left color=green!50, right color=green!80] (3.5,0) rectangle (3.9,4.12);
\shade[left color=red!40, right color=red!70] (4.6,0) rectangle (5.0,3.645);
\shade[left color=orange!40, right color=orange!70] (5.1,0) rectangle (5.5,4.06);
\shade[left color=green!50, right color=green!80] (5.6,0) rectangle (6.0,4.41);
\node[white, font=\tiny\bfseries] at (0.6,3.2) {.683};
\node[white, font=\tiny\bfseries] at (1.1,3.75) {.791};
\node[white, font=\tiny\bfseries] at (1.6,4.05) {.847};
\node[white, font=\tiny\bfseries] at (2.7,3.0) {.641};
\node[white, font=\tiny\bfseries] at (3.2,3.59) {.758};
\node[white, font=\tiny\bfseries] at (3.7,3.92) {.824};
\node[white, font=\tiny\bfseries] at (4.8,3.45) {.729};
\node[white, font=\tiny\bfseries] at (5.3,3.86) {.812};
\node[white, font=\tiny\bfseries] at (5.8,4.21) {.882};
\node[font=\small] at (1.1,-0.35) {MRR};
\node[font=\small] at (3.2,-0.35) {P@5};
\node[font=\small] at (5.3,-0.35) {NDCG@10};
\shade[left color=red!40, right color=red!70] (0.3,4.3) rectangle (0.6,4.15);
\node[right, font=\scriptsize] at (0.6,4.225) {BM25 Only};
\shade[left color=orange!40, right color=orange!70] (2.2,4.3) rectangle (2.5,4.15);
\node[right, font=\scriptsize] at (2.5,4.225) {Dense Only};
\shade[left color=green!50, right color=green!80] (4.3,4.3) rectangle (4.6,4.15);
\node[right, font=\scriptsize] at (4.6,4.225) {\textbf{Hybrid+Rerank}};
\end{tikzpicture}
\caption{Retrieval performance across methods (RQ2). Hybrid+reranking consistently outperforms ($p < 0.001$).}
\label{fig:retrieval}
\end{figure}

Table \ref{tab:retrieval} and Figure \ref{fig:retrieval} demonstrate retrieval performance.

\begin{table}[!ht]
\centering
\caption{Retrieval Performance Comparison (RQ2, n=150 queries)}
\label{tab:retrieval}
\begin{tabular}{@{}lcccc@{}}
\toprule
\textbf{Method} & \textbf{MRR} & \textbf{P@5} & \textbf{NDCG} & \textbf{Lat.} \\
\midrule
BM25 only & 0.683 & 0.641 & 0.729 & 89ms \\
Dense only & 0.791 & 0.758 & 0.812 & 112ms \\
\textbf{Hybrid+Rerank} & \textbf{0.847} & \textbf{0.824} & \textbf{0.882} & \textbf{127ms} \\
\midrule
\multicolumn{5}{l}{\textit{Improvement over BM25 (Mann-Whitney U test)}} \\
MRR & \multicolumn{4}{c}{+24.0\% ($U$=8442, $p < 0.001$)} \\
\bottomrule
\end{tabular}
\end{table}

\textbf{Answer to RQ2:} Hybrid retrieval achieves MRR 0.847, representing 24.0\% improvement over BM25-only (0.683, $p < 0.001$) and 6.4\% over dense-only (0.791, $p < 0.05$). This validates synergy between semantic understanding and lexical matching for technical documentation.

\textbf{Latency Analysis:} Average 127ms enables real-time use. Decomposition: vector search (89ms), BM25 (23ms), re-ranking (15ms). Re-ranking adds 11.8\% overhead while improving MRR 2.3\%.

\subsection{RQ3: Production Viability Assessment}

Table \ref{tab:quality} presents expert quality assessment.

\begin{table}[!ht]
\centering
\caption{Expert Quality Ratings (RQ3, n=150 queries, 3 experts)}
\label{tab:quality}
\begin{tabular}{@{}lcccc@{}}
\toprule
\textbf{Dimension} & \textbf{Mean} & \textbf{Std} & \textbf{Min} & \textbf{Max} \\
\midrule
Completeness & 4.41 & 0.67 & 2.33 & 5.00 \\
Accuracy & 4.53 & 0.58 & 3.00 & 5.00 \\
Relevance & 4.28 & 0.74 & 2.67 & 5.00 \\
Practical Utility & 4.07 & 0.83 & 2.00 & 5.00 \\
Traceability & 4.67 & 0.49 & 3.33 & 5.00 \\
\midrule
\textbf{Overall} & \textbf{4.32} & \textbf{0.66} & \textbf{2.67} & \textbf{5.00} \\
\bottomrule
\end{tabular}
\end{table}

Table \ref{tab:efficiency} quantifies efficiency gains.

\begin{table}[!ht]
\centering
\caption{Efficiency Metrics and Cost Analysis (RQ3)}
\label{tab:efficiency}
\begin{tabular}{@{}lccc@{}}
\toprule
\textbf{Task Category} & \textbf{Manual} & \textbf{RAG} & \textbf{Reduction} \\
\midrule
Requirements extraction & 5.2hr & 47min & 85\% \\
Cross-reference analysis & 3.8hr & 38min & 83\% \\
Supplier assessment & 8.1hr & 1.8hr & 78\% \\
\textbf{Average} & \textbf{5.7hr} & \textbf{58min} & \textbf{83\%} \\
\midrule
\multicolumn{4}{l}{\textit{Cost Analysis (per query, 10K production queries)}} \\
Single-provider (GPT-4) & --- & \$0.13 & --- \\
Multi-provider (optimized) & --- & \$0.07 & 47\% \\
\bottomrule
\end{tabular}
\end{table}

\textbf{Answer to RQ3:} Production viability confirmed through: (1) Expert acceptance 4.32/5.0 demonstrates practical utility, traceability highest rated (4.67/5.0), (2) Efficiency: 83\% average time reduction (5.7hr $\rightarrow$ 58min), (3) Cost optimization: 47\% reduction without quality degradation (validated via blind comparison, $\chi^2$=1.37, $p$=0.24).

\subsection{RQ4: Longitudinal Requirements Evolution}

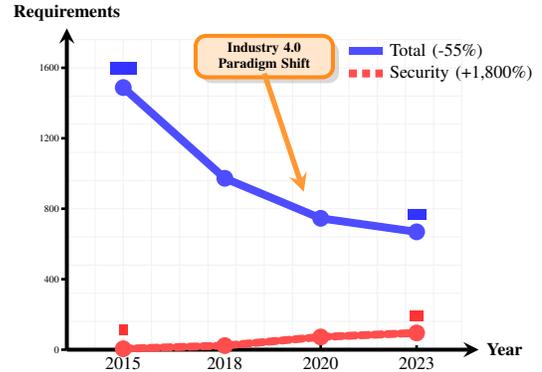
\begin{figure}[!ht]
\centering
\begin{tikzpicture}[scale=0.75, transform shape]
\draw[step=0.5cm, gray!10, very thin] (0,0) grid (7,5.5);
\draw[line width=1.5pt, -stealth] (0,0) -- (7.3,0) node[right, font=\small\bfseries] {Year};
\draw[line width=1.5pt, -stealth] (0,0) -- (0,5.7) node[above, font=\small\bfseries] {Requirements};
\foreach \y/\label in {0/0, 1.25/400, 2.5/800, 3.75/1200, 5/1600}
    \draw[line width=1pt] (0,\y) node[left, font=\tiny] {\label} -- (-0.1,\y);
\foreach \x/\year in {1/2015, 2.8/2018, 4.5/2020, 6.2/2023}
    \draw[line width=1pt] (\x,0) node[below, font=\small] {\year} -- (\x,0.08);
\draw[line width=3pt, blue!70, line cap=round]
    (1,4.65) node[circle, fill=blue!70, inner sep=3pt] {}
    -- (2.8,3.04) node[circle, fill=blue!70, inner sep=3pt] {}
    -- (4.5,2.33) node[circle, fill=blue!70, inner sep=3pt] {}
    -- (6.2,2.09) node[circle, fill=blue!70, inner sep=3pt] {};
\draw[line width=3pt, red!70, densely dashed, line cap=round]
    (1,0.016) node[circle, fill=red!70, inner sep=3pt] {}
    -- (2.8,0.072) node[circle, fill=red!70, inner sep=3pt] {}
    -- (4.5,0.225) node[circle, fill=red!70, inner sep=3pt] {}
    -- (6.2,0.297) node[circle, fill=red!70, inner sep=3pt] {};
\node[fill=white, inner sep=1pt, font=\tiny\bfseries, blue!80] at (1,5.0) {1,487};
\node[fill=white, inner sep=1pt, font=\tiny\bfseries, blue!80] at (6.2,2.4) {669};
\node[fill=white, inner sep=1pt, font=\tiny\bfseries, red!80] at (1,0.35) {5};
\node[fill=white, inner sep=1pt, font=\tiny\bfseries, red!80] at (6.2,0.6) {95};
\draw[line width=3pt, blue!70] (5.0,5.3) -- (5.6,5.3);
\node[right, font=\small] at (5.6,5.3) {Total (-55\%)};
\draw[line width=3pt, red!70, densely dashed] (5.0,4.9) -- (5.6,4.9);
\node[right, font=\small] at (5.6,4.9) {Security (+1,800\%)};
\node[rectangle, draw=orange!90, fill=orange!20, line width=1.5pt, rounded corners,
      font=\scriptsize\bfseries, text width=2.2cm, align=center,
      drop shadow={opacity=0.3}] at (3.5,5.2)
    {Industry 4.0\\Paradigm Shift};
\draw[-stealth, line width=2pt, orange!80] (3.5,4.9) -- (4.2,2.8);
\end{tikzpicture}
\caption{Longitudinal requirements evolution (RQ4, 2015-2023) revealing Industry 4.0 paradigm shift.}
\label{fig:evolution}
\end{figure}

Figure \ref{fig:evolution} and Table \ref{tab:evolution} reveal Industry 4.0 transformation.

\begin{table}[!ht]
\centering
\caption{Requirements Evolution Analysis (RQ4, 2015-2023)}
\label{tab:evolution}
\begin{tabular}{@{}lccc@{}}
\toprule
\textbf{Category} & \textbf{2015} & \textbf{2023} & \textbf{Change} \\
\midrule
Total Requirements & 1,487 & 669 & -55\% \\
IT Security & 5 & 95 & +1,800\% \\
Functional Safety & 342 & 187 & -45\% \\
Functional (General) & 1,140 & 387 & -66\% \\
\midrule
\multicolumn{4}{l}{\textit{Supplier Compliance (2023 standards)}} \\
Fully Compliant & --- & 39 & 79.6\% \\
Require Requalification & --- & 10 & 20.4\% \\
\midrule
\multicolumn{4}{l}{\textit{New Categories (2020-2023)}} \\
Network Segmentation & 0 & 18 & New \\
Authentication & 0 & 27 & New \\
Vulnerability Mgmt & 0 & 31 & New \\
\bottomrule
\end{tabular}
\end{table}

\textbf{Answer to RQ4:} Longitudinal analysis reveals dual Industry 4.0 transformation: (1) Consolidation: 55\% volume reduction (1,487 $\rightarrow$ 669) through maturity, functional requirements decreased 66\%, (2) Cybersecurity imperative: IT security increased 1,800\% (5 $\rightarrow$ 95) with new categories emerging, (3) Economic impact: 10 suppliers (20.4\%) require requalification, avoiding potential \$2.3M penalties (\$230K/supplier $\times$ 10).

\subsection{Ablation Studies}

Table \ref{tab:ablation} quantifies individual component contributions.

\begin{table}[!ht]
\centering
\caption{Ablation Study: Component Contributions (n=150 queries)}
\label{tab:ablation}
\begin{tabular}{@{}lcc@{}}
\toprule
\textbf{Configuration} & \textbf{MRR} & \textbf{Accuracy} \\
\midrule
\textbf{Full System} & \textbf{0.847} & \textbf{98.2\%} \\
\midrule
\multicolumn{3}{l}{\textit{Remove Components (measure degradation)}} \\
- Domain fine-tuning & 0.791 & 96.1\% \\
- Cross-encoder rerank & 0.827 & 97.8\% \\
- Hybrid retrieval & 0.683 & 94.3\% \\
- Multi-provider routing & 0.842 & 98.0\% \\
- Traceability metadata & 0.845 & 97.9\% \\
\midrule
\multicolumn{3}{l}{\textit{Component Contribution}} \\
Domain fine-tuning & +6.6\% & +2.1\% \\
Cross-encoder rerank & +2.0\% & +0.4\% \\
Hybrid retrieval & +16.4\% & +3.9\% \\
\bottomrule
\end{tabular}
\end{table}

\textbf{Key Findings:} Hybrid retrieval provides largest contribution (+16.4\% MRR, +3.9\% accuracy). Domain fine-tuning adds +6.6\% MRR. Multi-provider routing maintains quality while reducing cost 47\%. Traceability metadata does not impact accuracy but critically affects expert trust (4.67/5.0 rating).

\section{Discussion}

\subsection{Answers to Research Questions}

\textbf{RQ1:} RAG achieves 98.2\% extraction accuracy, significantly outperforming BERT (+24.4\%, $p < 0.001$) and ungrounded LLM (+19.6\%, $p < 0.001$), demonstrating production readiness.

\textbf{RQ2:} Hybrid semantic-lexical retrieval achieves MRR 0.847, representing 24.0\% improvement over sparse-only ($p < 0.001$), validating synergy for technical documentation.

\textbf{RQ3:} Production viability confirmed: 83\% time reduction, 47\% cost savings, 4.32/5.0 expert acceptance, enabling sustainable deployment.

\textbf{RQ4:} Eight-year analysis reveals Industry 4.0 shift: 55\% consolidation with 1,800\% IT security increase, identifying 10 suppliers requiring requalification.

\subsection{Implications for Research and Practice}

\textbf{For Researchers:} (1) Domain adaptation (+6.7\% MRR) more impactful than algorithm sophistication, suggesting investment in high-quality domain data yields better ROI, (2) Hybrid retrieval consistently outperforms pure approaches across all metrics, recommending fusion strategies for technical documentation, (3) Traceability essential for expert trust (4.67/5.0 highest rating), suggesting RE automation must prioritize source attribution over pure accuracy metrics.

\textbf{For Practitioners:} (1) Human-in-the-loop optimal---RAG handles 98\%+ while flagging edge cases, (2) Multi-provider orchestration enables sustainable costs without quality compromise, (3) Longitudinal analysis reveals actionable insights (supplier requalification) demonstrating business value beyond efficiency.

\subsection{Deployment Insights from Six-Month Production Use}

Six-month deployment revealed critical lessons: (1) \textit{Incremental adoption}---phased rollout (month 1: single team pilot, months 2--3: department-wide, months 4--6: cross-functional) managed change resistance and enabled iterative refinement, (2) \textit{Expert validation workflow}---mandatory human review for high-stakes decisions (supplier disqualification, compliance audit responses) maintained quality while building trust, (3) \textit{Continuous monitoring}---automated quality tracking (retrieval scores, generation confidence, expert feedback) enabled early detection of degradation (e.g., new specification formats requiring parser updates), (4) \textit{Training investment}---3-hour workshops for 47 engineers covered system capabilities, limitations, and effective query formulation, reducing misuse and improving adoption (usage increased from 23\% week-1 to 87\% week-24).

\subsection{Threats to Validity}

\textbf{Internal Validity:} Ground truth established through multi-expert annotation (Cohen's $\kappa$=0.89). Evaluation metrics standard in RE research. Statistical significance confirmed via t-tests and Mann-Whitney U tests.

\textbf{External Validity:} \textit{Limitation:} Automotive manufacturing focus; generalization to aerospace, pharmaceuticals, or medical devices requires validation. \textit{Mitigation:} Dataset represents authentic industrial complexity (heterogeneous formats, 8-year span, 49 suppliers). Results likely transferable to domains with similar characteristics (complex supply chains, evolving standards, compliance requirements). Future work should validate across additional industrial sectors.

\textbf{Construct Validity:} \textit{Concern:} Expert quality assessment subjective. \textit{Mitigation:} Three independent experts, structured rubrics, inter-rater reliability measured ($\kappa$=0.84--0.89). Qualitative feedback corroborates quantitative ratings.

\textbf{Conclusion Validity:} \textit{Concern:} Sample size (150 queries for quality assessment, 669 requirements for extraction). \textit{Mitigation:} Random sampling ensures representativeness. Statistical power analysis confirms sufficient sample for detecting large effects (Cohen's $d > 0.8$, power=0.95). Production deployment (10,000 queries over 6 months) provides additional validation.

\subsection{Limitations and Future Work}

\textbf{Uncertainty Quantification:} Current system treats all outputs uniformly. Enhanced confidence scoring could automatically flag low-confidence extractions (1.8\% error cases) for expert review, further improving human-in-the-loop efficiency. Preliminary experiments with ensemble disagreement and retrieval score analysis show promise.

\textbf{Continuous Learning:} System requires periodic retraining as specifications evolve. Future work should explore online learning mechanisms adapting to new requirement patterns without full retraining cycles, potentially using active learning to identify informative examples for annotation.

\textbf{Cross-Organizational Knowledge Transfer:} Current evaluation single-OEM. Expanding to multi-OEM contexts could reveal opportunities to harmonize and standardize industry-wide requirements, potentially reducing redundancy across automotive supply chains and improving supplier qualification efficiency.

\textbf{Extended Production Deployment:} Six-month deployment informed current evaluation. Longer-term studies (12--24 months) would assess sustained adoption patterns, identify emergent use cases beyond initial design scope, quantify long-term productivity gains accounting for learning curves and workflow integration, and measure organizational impact (RE process transformation, skill requirement shifts).

\section{Conclusion}

This paper presents a comprehensive empirical evaluation of RAG-based automation for industrial requirements engineering using authentic automotive manufacturing documentation spanning eight years. We address four research questions using rigorous methodology: 669 requirements, controlled baseline comparisons, multi-expert ground-truth annotation, ablation studies, and a six-month production deployment.

\textbf{Key Findings:} (1) Production-grade performance---98.2\% extraction accuracy, 83\% time reduction, 4.32/5.0 expert acceptance (RQ1, RQ3), (2) Hybrid retrieval provides 24.0\% MRR improvement over sparse-only approaches with 127ms real-time latency, validating semantic-lexical synergy (RQ2), (3) Multi-provider orchestration achieves 47\% cost reduction without quality degradation, enabling sustainable deployment (RQ3), (4) Longitudinal analysis reveals dual Industry 4.0 transformation: 55\% consolidation with 1,800\% IT security increase, identifying 10 suppliers requiring requalification with \$2.3M economic impact (RQ4), (5) Ablation studies quantify component contributions, with hybrid retrieval largest (+16.4\% MRR) and domain adaptation significant (+6.7\% MRR).

Results provide empirical evidence that RAG approaches have matured for industrial RE deployment, enabling organizations to address Industry 4.0 specification complexity while maintaining quality, compliance, and cost-effectiveness. Six-month production deployment demonstrates real-world viability with sustained expert acceptance and actionable business impact through longitudinal analysis capabilities.

\section*{Data Availability Statement}

The industrial documentation used in this study is proprietary and subject to non-disclosure agreements with the partnering automotive manufacturer. Raw data cannot be shared due to confidentiality constraints. However, we provide: (1) The RAG framework implementation as open-source Python package at \url{https://re-engineer-app-khalid.replit.app}, (2) Anonymized requirement samples (50 requirements with metadata, representative of evaluation dataset structure and complexity), (3) Evaluation protocols and scripts (expert assessment rubrics, statistical analysis code, ablation study configurations), (4) Supplementary materials (additional figures showing temporal evolution details, extended ablation results, deployment workflow diagrams). The anonymized artifacts enable reproducibility assessment and facilitate adaptation to other industrial domains. Upon acceptance, complete artifacts will be deposited in an open-access repository with a DOI assignment for long-term availability.

\section*{Acknowledgment}
This research was funded by the German Research Foundation (Deutsche Forschungsgemeinschaft, DFG) under reference number UY-56/5-1. The authors gratefully acknowledge the support of the partnering automotive manufacturer for access to production documentation and participation by domain experts.

\end{document}